\documentclass[prl,aps,superscriptaddress,
                           noshowpacs,twocolumn,reprint,noshowkeys,
                           amssymb,amsmath,a4paper]{revtex4-1}
\usepackage{microtype} %For better kerning and symbol-stretching
\usepackage{xspace} %For the \xspace command

\usepackage{txfonts}  %Times Roman fonts
\usepackage{txfontsb} %Addition for txfonts, including old style numerals and greek

\usepackage{bm} %Bold math symbols with \bm{} (Greek and other symbols)
\usepackage{moresize} %Enables the use of \ssmall which lies between \tiny and \scriptsize

\usepackage{xcolor}
\usepackage{graphicx}

\usepackage{hyperref}
\hypersetup{colorlinks,
	        linkcolor={blue!75!black!80!yellow},
    	    citecolor={blue!75!black!80!yellow},
        	urlcolor={blue!75!black!80!yellow}
			}

\usepackage{placeins}

\usepackage{siunitx}
\usepackage[centering,hmargin=16mm,tmargin=30mm,bmargin=26mm]{geometry}

\newcommand{\SSS}{\scriptscriptstyle}

\newcommand{\Dd}{{\rm d}}
\newcommand{\Ii}{{\rm i}}

\newcommand{\omegac}{\omega_{\text{c}}}
\newcommand{\omegaf}{\omega_{\text{\textsc{f}}}}
\newcommand{\omegak}{\omega_{\mathbf{K}}}
\newcommand{\vf}{v_{\SSS\mathrm{F}}}
\newcommand{\ef}{E_{\SSS\mathrm{F}}}

\newcommand{\rv}{\mathbf{r}}
\newcommand{\pv}{\mathbf{p}}
\newcommand{\kv}{\mathbf{k}}
\newcommand{\Jv}{\mathbf{J}}

\newcommand{\ie}{i.e.\@\xspace} %Gobble-spaces of the "small" type (small is ensured by adding \@)
\newcommand{\cf}{cf.\@\xspace}
\newcommand{\eg}{e.g.\@\xspace}

\newcommand{\appropto}{\mathrel{\vcenter{
			\offinterlineskip\halign{\hfil$##$\cr
				\propto\cr\noalign{\kern.2pt}\sim\cr\noalign{\kern-2.5pt}}}}}
			
\DeclareMathOperator{\sgn}{sgn}
\let\Re\relax %Remove the default definition before redefining
\DeclareMathOperator{\Re}{Re}
\let\Im\relax %Remove the default definition before redefining
\DeclareMathOperator{\Im}{Im}

\begin{document}

%-----------------
%----- TITLE -----
%-----------------
\title{Infrared Topological Plasmons in Graphene}

%------------------------------------
%----- AUTHORS AND AFFILIATIONS -----
%------------------------------------
\author{Dafei~Jin} %
\email{dafeijin@berkeley.edu}\email{tchr@mit.edu}\thanks{these authors contributed equally to this work.}
\affiliation{Department of Mechanical Engineering, University of California, Berkeley, California 94720, USA}
\author{Thomas~Christensen}
\email{dafeijin@berkeley.edu}\email{tchr@mit.edu}\thanks{these authors contributed equally to this work.}
\affiliation{Department of Physics, Massachusetts Institute of Technology, Cambridge, Massachusetts 02139, USA}
\author{Marin~Solja\v{c}i\'{c}}
\affiliation{Department of Physics, Massachusetts Institute of Technology, Cambridge, Massachusetts 02139, USA}
\author{Nicholas~X.~Fang}
\affiliation{Department of Mechanical Engineering, Massachusetts Institute of Technology, Cambridge, Massachusetts 02139, USA}
\author{Ling~Lu} \email{linglu@iphy.ac.cn}
\affiliation{Institute of Physics, Chinese Academy of Sciences/Beijing National Laboratory for Condensed Matter Physics, Beijing 100190, China}
\author{Xiang~Zhang} \email{xzhang@me.berkeley.edu}
\affiliation{Department of Mechanical Engineering, University of California, Berkeley, California 94720, USA}

%--------------------
%----- ABSTRACT -----
%--------------------
\begin{abstract}
We propose a two-dimensional plasmonic platform -- periodically patterned monolayer graphene -- which hosts topological one-way edge states operable up to infrared frequencies. We classify the band topology of this plasmonic system under time-reversal-symmetry breaking induced by a static magnetic field. At finite doping, the system supports topologically nontrivial bandgaps with mid-gap frequencies up to tens of terahertz. By the bulk-edge correspondence, these bandgaps host topologically protected one-way edge plasmons, which are immune to backscattering from structural defects and subject only to intrinsic material and radiation loss. Our findings reveal a promising approach to engineer topologically robust chiral plasmonic devices and demonstrate a realistic example of high-frequency topological edge state.
\end{abstract}

%---------------------
%----- MAKETITLE -----
%---------------------
\maketitle
\pretolerance=8000 

%---------------------
%----- MAIN TEXT -----
%---------------------
Time-reversal-symmetry $(\mathcal{T})$ breaking, a necessary condition for achieving quantum Hall phases~\cite{Klitzing1980IQHE,haldane1988model}, has now been successfully implemented in several bosonic systems, as illustrated by the experimental observation of topologically protected one-way edge transport of photons~\cite{Wang:Nature2009,Skirlo:2015} and phonons~\cite{Nash:2015}.
More generally, two-dimensional (2D) $\mathcal{T}$-broken topological bosonic phases have been proposed in a range of bosonic phases, spanning photons~\cite{Haldane:2008}, phonons~\cite{Prodan:2009, Wang:2015Topological}, magnons~\cite{Shindou:2013}, excitons~\cite{YuenZhou:2013}, and polaritons~\cite{Karzig:2015}.
The operating frequency of these systems is typically small, however -- far below \si{THz} -- limited by the spectral range of the $\mathcal{T}$-breaking mechanism. For example, the gyromagnetic effect employed in topological photonic crystals is limited by the Larmor frequency of the underlying ferrimagnetic resonance, on the order of tens of \si{GHz}~\cite{Wang:Nature2009}.
In phononic realizations, the attainable gyrational frequencies limit operation further still, to the range of \si{kHz}~\cite{fleury2014sound}.
Towards optical frequencies, proposals of dynamic index modulation~\cite{fang2012realizing} and optomechanical coupling~\cite{hafezi2012optomechanically} are promising
but experimentally challenging to scale to multiple coupled elements~\cite{tzuang2014non, shen2016experimental, fang2016generalized}.%

%--------------------------------------------------------------------------------
\begin{figure}[!tb]
	\centerline{\includegraphics[scale=1]{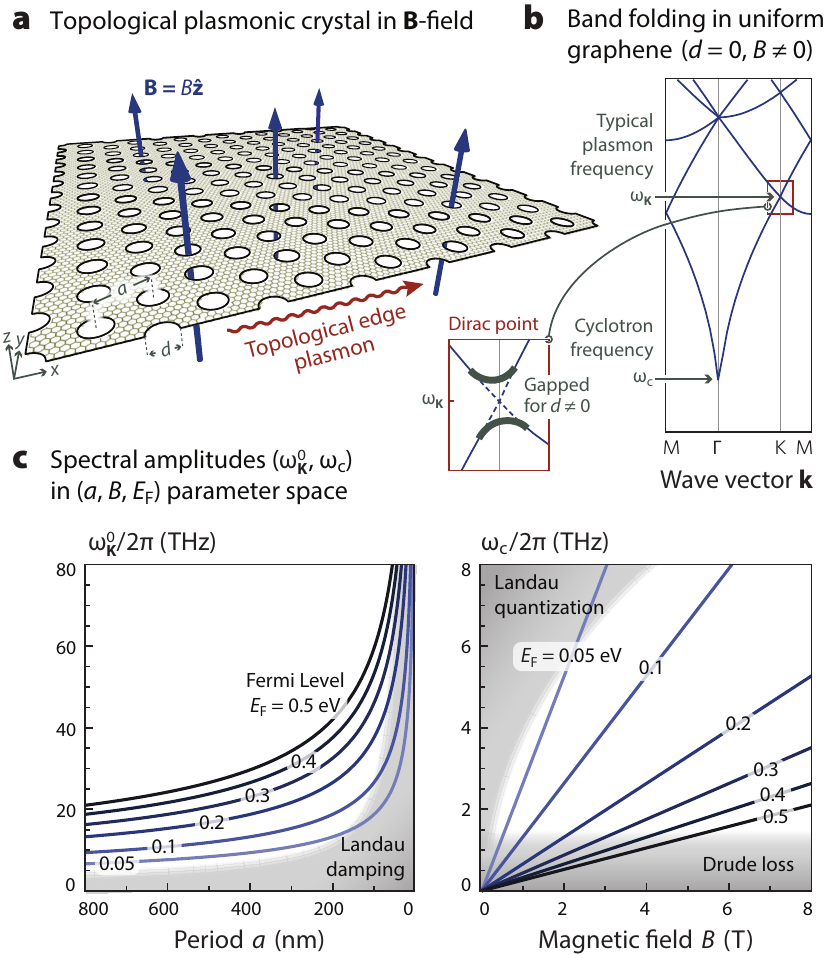}}
	\caption{2D topological plasmonic crystal under magnetically induced $\mathcal{T}$-breaking.
		(a)~Schematic of triangular antidot lattice in graphene. Under an external magnetic field $\mathbf{B}=B\hat{\mathbf{z}}$, a finite lattice supports topologically protected one-way edge plasmons.
		(b)~Band-folded plasmon-dispersion in uniform graphene at $B\neq 0$; characteristic frequencies $\omegak$ and $\omegac$ indicated. The symmetry-induced Dirac cone is gapped for $d\neq 0$.
		(c)~Characteristic frequencies' dependence on the crystal period $a$, magnetic field $B$, and Fermi level $\ef$.
	}
	\label{fig:generalproperties}%
\end{figure}
%--------------------------------------------------------------------------------

Recently, \citet{Jin:2016} pointed out that the well-known magnetoplasmons of uniform 2D electron gases~\cite{Mast:1985,Volkov:1988} constitute an example of a topologically nontrivial bosonic phase hosting unidirectional edge states.
However, as the topological gap exists only below the cyclotron frequency $\omegac$, the spectral operation remains limited to low frequencies.
In this letter, we show that by suitably engineering the plasmonic band structure of a periodically nanostructured 2D monolayer graphene, see Fig.~\ref{fig:generalproperties}(a), the operation frequency of topological plasmons can be raised dramatically, to tens of \si{THz}, while maintaining large gap--midgap ratios even under modest $\mathbf{B}$-fields. Bridging ultrafast electronics and infrared topological photonics, the proposed platform can be seamlessly integrated with well-established CMOS technology, allowing dynamically gate-tunable topological states across a broad spectral range.

Graphene distinguishes itself as an ideal platform for topological plasmonics in three key aspects: first, it supports large, tunable carrier densities $n\!\sim\! \SIrange{e11}{e14}{\per\cm\squared}$~\cite{Efetov:2010,Liu:2011,Fang:2014}, or equivalently, large, tunable Fermi energies $\ef = \hbar\vf\sqrt{\pi n}$ (Fermi velocity, $\vf\approx \SI{9.1e7}{\cm\per\s}$~\cite{Note1}); second, it exhibits an ultrasmall, tunable Drude mass $m^* \equiv \ef/\vf^2$ (\eg at $\ef = \SI{0.2}{\eV}$, $m^*/m_{\text{e}} \approx 4\%$), allowing ultrahigh cyclotron frequencies $\omegac \equiv eB/cm^* = eB\vf^2/c\ef$ up to the THz range; and third, high-quality graphene can exhibit exceptionally long intrinsic relaxation times $1/\gamma$, extending into the picosecond range~\cite{Du:2008,Bolotin:2008}.
These properties enable topological plasmons of unprecedentedly high frequency, short wavelength, long propagation, and large topological bandgaps.

%----------------------------------------------------------------------
\begin{figure*}[!htb]
	\centerline{\includegraphics[scale=1]{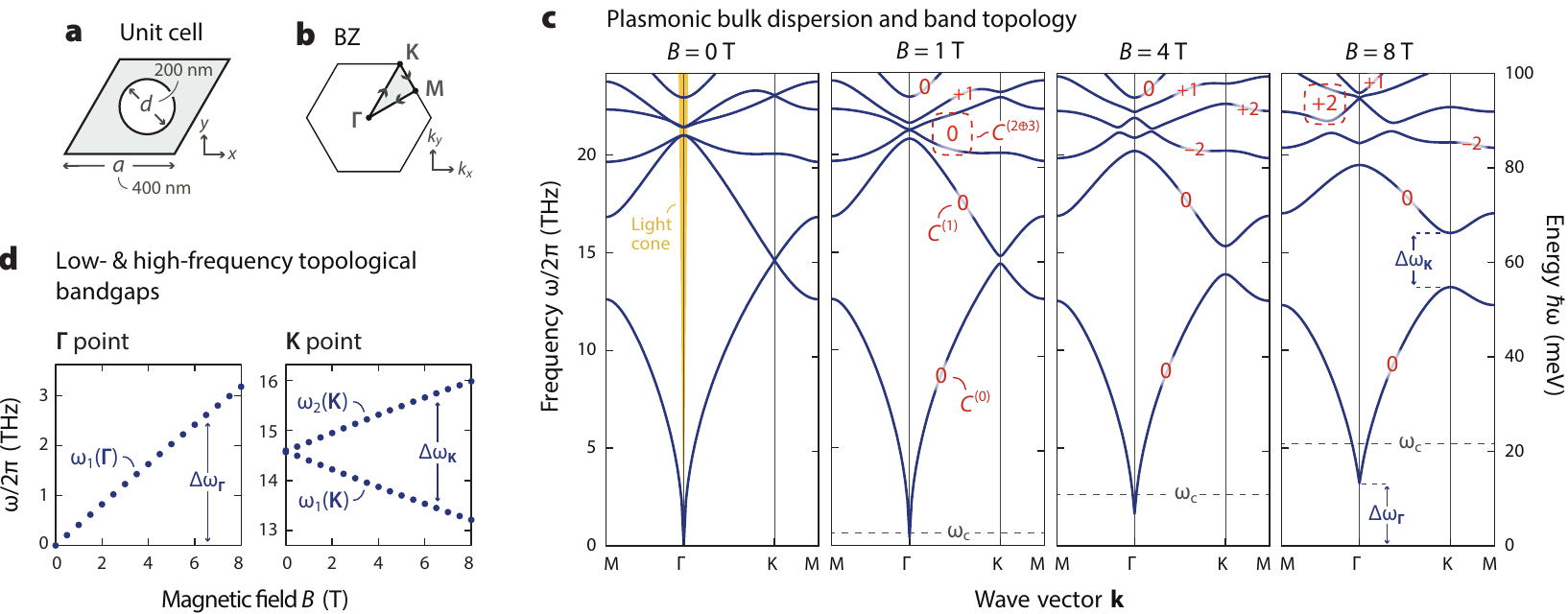}}
	\caption{Bulk properties.
		(a)~Unit cell.
		(b)~Brillouin zone.
		(c)~Bulk dispersion along the high-symmetry directions of the irreducible BZ for $B=\SIlist{0; 1; 4; 8}{\tesla}$. Chern numbers are indicated in orange labels; composite Chern numbers are highlighted by a dashed periphery.
		(d)~Splitting of $\boldsymbol{\Gamma}$ and $\mathbf{K}$ point degeneracies and opening of low- and high-frequency topological bandgaps with increasing magnetic field.
	}
	\label{fig:bulkproperties}%
\end{figure*}
%----------------------------------------------------------------------

The plasmonic properties of a general graphene domain $\rv \in \Omega\subseteq \mathbb{R}^2$ under an external magnetic field $\mathbf{B} = B\hat{\mathbf{z}}$ is described by a linear eigenvalue problem with three field components: the scalar potential $\Phi$ and the surface electric current density $\mathbf{J}\equiv \mathrm{J}_x\hat{\mathbf{x}}+\mathrm{J}_y\hat{\mathbf{y}}$~\cite{Jin:2016}.
For an eigenstate indexed by $\nu$ and frequency $\omega_\nu$, this eigenproblem is specified by~\cite{Note2}%
\begin{subequations}\label{eqs:governing}
	\begin{align}
		&\hat{\mathcal{H}} \mathbf{U}_\nu = \omega_\nu \mathbf{U}_\nu ,
		\label{eq:governing_form}\\
		&\text{with}\ \mathbf{U}_\nu \equiv \begin{pmatrix} \omegaf\Phi \\ \mathbf{J} \end{pmatrix}\
		\text{and}\		\hat{\mathcal{H}} \equiv \begin{pmatrix} 0 & \omegaf\hat{V} \hat{\mathbf{p}}^{\text{T}}  \\ \alpha \hat{\mathbf{p}} & \omegac\sigma_2 \end{pmatrix}.
		\label{eq:governing_explicit}	
	\end{align}
\end{subequations}
Here, $\hat{\pv}\equiv -\Ii\bm{\nabla}$ is the in-plane momentum operator, $\hat{V}[f](\rv)\equiv\int_{\Omega} \Dd\rv'\, f(\rv') / |\rv-\rv'|$ the Coulomb operator, $\sigma_2\equiv \Big(\begin{smallmatrix} 0 & -\Ii\\ \Ii & 0\end{smallmatrix}\Big)$ a Pauli matrix, $\omegaf\equiv \ef/\hbar$ the Fermi ``frequency'', and $\alpha\equiv e^2/\pi\hbar$ a prefactor of graphene's intraband conductivity $\alpha\omegaf\omega^{-1}$. Conceptually, Eqs.~\eqref{eqs:governing} and comprise the Coulomb, continuity, and constitutive equations.
The no-spill boundary condition $\Jv\cdot\hat{\mathbf{n}}=0$ applies along the perimeter of $\Omega$ (edge normal, $\hat{\mathbf{n}}$). Under a suitable inner product Eq.~\eqref{eq:governing_form} is Hermitian (see SM).

We explore the band topology of 2D plasmons in periodically structured graphene under magnetic-field induced $\mathcal{T}$-breaking.
Figure \ref{fig:generalproperties}(a) illustrates our design: a triangular antidot lattice of periodicity $a$ and antidot diameter $d$ is etched into a suspended sheet of graphene~\cite{Note3}.
The domain $\Omega$ in Eqs.~\eqref{eqs:governing} is then the torus defined by the rhombic unit cell of Fig.~\ref{fig:bulkproperties}(a). Band folding splits the eigenindex $\nu$ into a band index $n = 1,2,\ldots$ and a crystal wave vector $\mathbf{k}$ restricted to the hexagonal Brillouin zone (BZ) of Fig.~\ref{fig:bulkproperties}(b). Accordingly, the eigenvectors assume the Bloch form $\mathbf{U}_{n\mathbf{k}}(\rv) = \mathbf{u}_{n\mathbf{k}}(\rv)\mathrm{e}^{\Ii \mathbf{k}\cdot\mathbf{r}}$, with periodic component $\mathbf{u}_{n\mathbf{k}} \equiv (\omegaf\phi, \mathbf{j})_{n\mathbf{k}}^{\text{T}}$.

First, we consider the simple but instructive $d=0$ scenario, \ie the uniform sheet, see Fig.~\ref{fig:generalproperties}(b). This ``empty lattice'' captures the essential impact of band-folding: by folding the uniform sheet plasmon dispersion, $\omega(k) = \sqrt{2\pi\alpha\omegaf k+\omegac^2}$~\cite{Note4}, over the hexagonal BZ, three-fold Dirac-like point degeneracies arise between the $n=\numlist{1;2;3}$ bands at the $\mathbf{K}$ (and $\mathbf{K}'$) point.  For $B=0$, the lattice's $C_{6v}$ symmetry guarantees that two-fold degenerate Dirac points remain between the $n=\numlist{1;2}$ bands even when $d\neq 0$.
The uniform-sheet Dirac point plasmon frequency,
$\omegak \equiv \sqrt{(\omegak^0)^2+\omegac^2}$ with $\omegak^0 \equiv \sqrt{2\pi\alpha\omegaf|\mathbf{K}| }$ and $|\mathbf{K}|=4\pi/3a$,
along with the cyclotron frequency $\omegac$, then define the characteristic frequencies of the problem and are indicated in Fig.~\ref{fig:generalproperties}(b). By applying a finite $B$-field to the $d\neq 0$ system, the Dirac point degeneracy is split, inducing a gap linearly proportional to $\omegac$. As a result, topological plasmons with both high frequency and sufficient topological gap require simultaneously large $\omegak^{(0)}$ and $\omegac$.

The parameter space involved in simultaneously maximizing $\omegak^0$ and $\omegac$ is illustrated in Fig.~\ref{fig:generalproperties}(c). The monotonic $\ef$-dependence of the two characteristic frequencies is opposite, highlighting an inherent trade-off between the operating frequency and the gap size. In addition, the accessible parameter space is restricted by several constraints, indicated by gray regions in Fig.~\ref{fig:generalproperties}(c): first, intrinsic Drude loss estimated at $\gamma/2\pi \sim \SI{1}{THz}$ smears out the gap region, necessitating $\omegac \gtrsim \gamma$; second, interband dispersion is non-negligible when $\omegak \gtrsim \omegaf$, eventually introducing significant loss through Landau damping; and third, Landau quantization of the charge carriers ultimately invalidates a semiclassical description~\cite{Gusynin:2007,Ferreira:2012} when $\ef\lesssim E_{\text{\textsc{l}}} \equiv \vf\sqrt{2\hbar e B /c}$ (the first Landau level), or equivalently, when $\hbar\omegac\lesssim \tfrac{1}{2} E_{\text{\textsc{l}}}$, see Supplemental Material (SM). Overall, we find that an experimentally favorable region exists for Fermi energies $\ef\sim\SIrange{0.2}{0.3}{\eV}$, periodicities $a\sim \SIrange{400}{600}{\nm}$, and magnetic fields $B\sim\SIrange{2}{8}{\tesla}$.

%--------------------------------------------------------------------------------
\begin{figure*}[htb]
	\centerline{\includegraphics[scale=1]{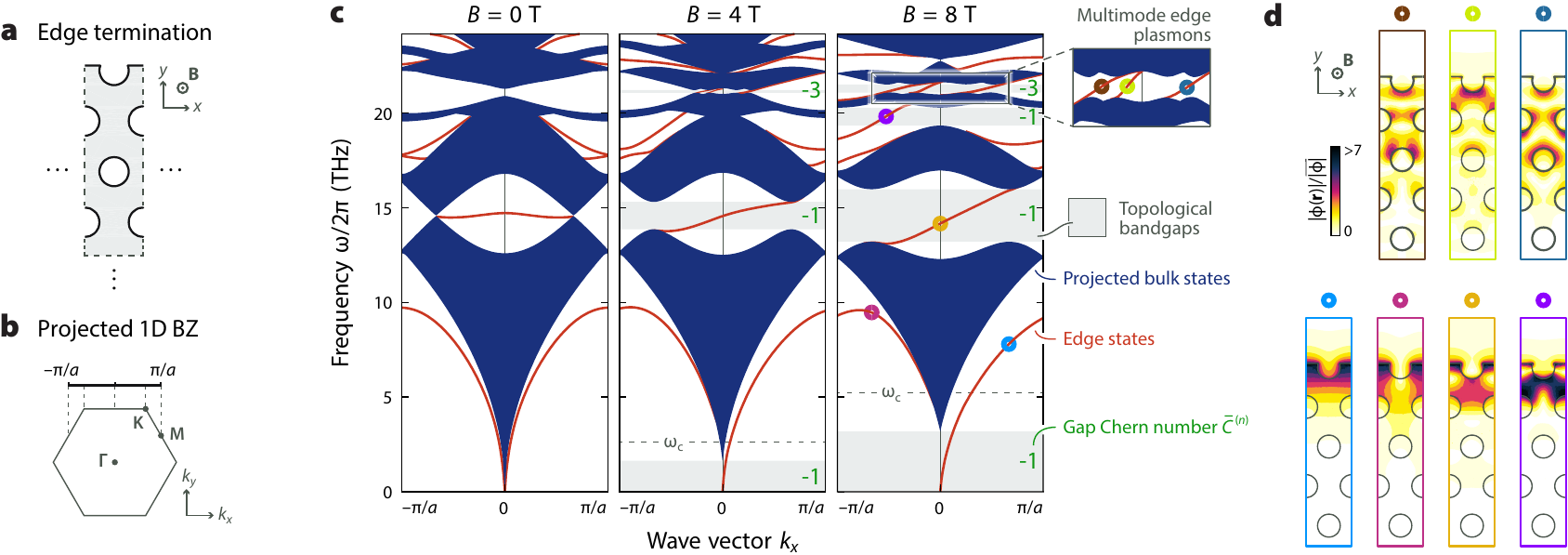}}
	\caption{Plasmonic one-way edge states at lattice terminations.
		(a)~Edge termination of the 2D crystal.
		(b)~Projected 1D BZ and its high symmetry points.
		(c)~Projected bulk bands (blue) and topologically protected one-way plasmonic edge states (red) along $k_x$ for $B=\SIlist{0; 4; 8}{\tesla}$, with associated gap Chern numbers $\bar{C}^{(n)}$ (green).
		(d)~Typical mode profiles of edge states in real space at $B=\SI{8}{\tesla}$; band-association is indicated by colored markers in (c).}
	\label{fig:edgestates}%
\end{figure*}
%--------------------------------------------------------------------------------

Next, we turn to the nanostructured system, settling on a periodicity $a=\SI{400}{\nm}$, antidot diameter $d=\SI{200}{\nm}$, see Fig.~\ref{fig:bulkproperties}(a), and a Fermi level $\ef = \SI{0.2}{\eV}$ (equivalent, at $B=0$, to a carrier density $n\approx \SI{3e12}{\per\cm\squared}$). Antidot lattices like these are well-within experimental capabilities~\cite{Bai:2010,Zhu:2014,Yeung:2014,Liu:2015}.
The eigenvalue problem, Eqs.~\eqref{eqs:governing}, is solved numerically by discretizing in an unstructured triangular mesh, employing linear nodal functions, and with the lattice-specific Coulomb interaction evaluated by Ewald summation (see SM).
Figure~\ref{fig:bulkproperties}(c) depicts the calculated plasmon dispersion $\omega_n(\kv)$ along the boundary of the irreducible BZ for increasing magnetic field strength $B=\SIlist{0; 1; 4; 8}{\tesla}$.

In the nonmagnetic scenario, $B=0$, the lattice disperses like the uniform sheet under the substitution $k\rightarrow \zeta_n(\kv)/a$, \ie as $\omega_n^0(\kv) = \sqrt{2\pi\alpha\omegaf\zeta(\kv)/a}$, with modal parameter $\zeta_n(\kv)$ solely dependent on $a/d$ and the \emph{relative} location of $\kv$ in the BZ~\cite{Christensen:2017}; \eg at $a/d=2$ we find $\zeta_{1,2}(\mathbf{K}) \approx 2.535$. Near the $\mathbf{\mathbf{\Gamma}}$ point $\zeta_1(\kv)\appropto |\mathbf{k}|$, yielding the conventional long-wavelength 2D plasmon dispersion $\omega\appropto\!\sqrt{k}$. Particle-hole symmetry ($\mathcal{C}$) of Eqs.~\eqref{eqs:governing} entails the existence of a corresponding set $\{n=-1,-2,\ldots\}$ of negative energy states, $\omega_{-n}(\kv) = -\omega_{n}(-\kv)$ (and a trivial zero-frequency band, $n=0$)~\cite{Jin:2016}: accordingly, besides the Dirac point degeneracy at $\mathbf{K}$ between the $n=1$ and $2$ bands, an implicit degeneracy exists at $\boldsymbol{\Gamma}$ between the $n=\pm 1$ (and $n=0$) bands.
By applying a magnetic field, the bands are linearly perturbed from $\omega_n^0(\kv)$ to $\omega_n(\kv) \simeq \omega_n^0(\kv) + \xi_n(\kv)\omegac+\mathcal{O}(\omegac^2)$ (see SM); the modal perturbation parameter $\xi_n(\kv)$ is obtained numerically at the degeneracy points as $\xi_1(\boldsymbol{\Gamma}) \approx 0.63$ and $\xi_{1,2}(\mathbf{K})\approx \mp 0.27$ at $a/d=2$~\cite{Note5}.
This is illustrated in Fig.~\ref{fig:bulkproperties}(d): the degeneracies at $\mathbf{\Gamma}$ and $\mathbf{K}$ are linearly and evenly gapped when $B\neq 0$. As we explain shortly, the low-frequency gap opened at $\boldsymbol{\Gamma}$ supports a topological edge state entirely analogous to its uniform sheet counterpart. The high-frequency (${\approx}\SI{15}{\THz}$) gap opening at $\mathbf{K}$, however, introduces a new, qualitatively distinct topological edge state.

Next, we describe the topological properties of the plasmonic lattice as quantified by the band Chern number, $C^{(n)} \equiv \frac{1}{2\pi\Ii} \oint_{\scriptscriptstyle \partial \text{BZ}} \Dd\kv\ \langle \mathbf{u}_{n\kv}|\nabla_{\kv}|\mathbf{u}_{n\kv} \rangle$ (evaluated numerically from the computed eigenvectors~\cite{Fukui:2005}). Figure~\ref{fig:bulkproperties}(b) depicts the evolution of $C^{(n)}$ across $B=\SIlist{0;1;4;8}{\tesla}$. At $B=\SI{0}{\tesla}$ the Berry flux is identically zero \cf time-reversal and parity symmetry; the band-structure is topologically trivial.
For $B\neq 0$, $\mathcal{T}$ is broken, allowing nonzero Berry fluxes and nontrivial topology: the 1st and 2nd bands have $C^{( 1 )}=C^{( 2 )}=0$, independent of $B$ since finite gaps separate them from distinct bands throughout the considered $B$-range. Conversely, the higher order bands, $n=3,4,\ldots$, display Chern numbers covering a broader range, up to $\pm 2$. A few bands exhibit point-degeneracies within numerical accuracy and are assigned a composite Chern number $C^{(n\oplus n+1)}$. As the $B$-field is increased, there is an exchange of Chern numbers between the $n=\numlist{4;5;6}$ bands as gaps close and reopen, illustrating the mechanism of Berry flux monopole exchange. For even stronger $B$-fields (see SM), all six bands eventually separate completely, leaving $C^{(1)} = C^{(2)} = 0$, $C^{(3)}=-2$, and $C^{( 4 )}=C^{( 5 )}=C^{( 6 )} = +1$.

%--------------------------------------------------------------------------------
\begin{figure}[htb]
\centerline{\includegraphics[scale=1]{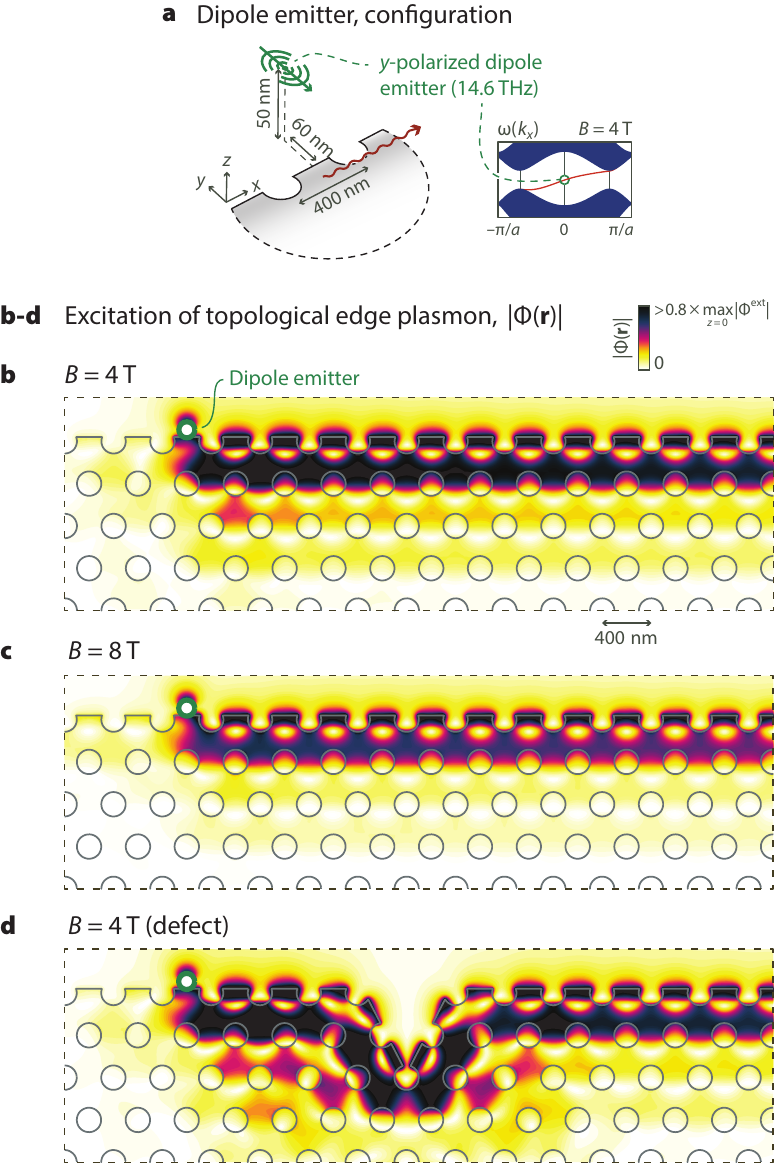}}
        \caption{Dipole excitation of edge plasmons in the $\bar{C}^{(2)} = -1$ gap.
        	(a)~Dipole configuration relative to nanostructured graphene edge.
        	(b--d)~Total potential $|\Phi(\rv)|$ of dipole-excited one-way edge plasmons at (b)~$B = \SI{4}{\tesla}$, (c)~$B=\SI{8}{\tesla}$, and (d)~$B=\SI{4}{\tesla}$ with a structural defect.}
\label{fig:dipoleexcitation}
\end{figure}
%--------------------------------------------------------------------------------

By the bulk-edge correspondence, the existence of topologically protected edge states is intimately linked with the bulk topology, \ie with $C^{(n)}$. As recently pointed out in Ref.~\citenum{Jin:2016}, the presence of $\mathcal{C}$ symmetry, and the concomitant existence of a set of negative-frequency states $\{n=-1,-2,\ldots\}$, necessitates a global perspective of the band topology for the definition of associated gap Chern numbers. Specifically, the total Chern number of positive ($+$) and negative ($-$) frequency bands is $C_\pm \equiv \sum_{n=1}^\infty C^{(\pm n)}$. In uniform graphene $C_\pm = \pm \sgn B$~\cite{Jin:2016}. Since Chern numbers can be annihiliated or created (pairwise) only under band closings, this result holds in nanostructured graphene as well \cf the finite bandgap separating positive and negative bands. With this in mind, we define the $n$th gap Chern number $\bar{C}_n$ associated with the gap immediately below the $n$th band as
\begin{equation}
\bar{C}^{( n )} \equiv \sum_{n'=-\infty}^{n-1}C^{(n')} = -\sgn B + \sum_{n'=1}^{n-1} C^{(n')} ,
\end{equation}
specializing to positive-frequency gaps at the last equality. For lattice terminations adjacent to vacuum, bulk-edge correspondence then requires that the number of left minus right propagating topological edge states equal $\bar{C}^{(n)}$~\cite{Hatsugai:1993}.

These considerations predict the existence of single-mode one-way edge states in the first and second gaps when $B\neq 0$ and multi-mode one-way edge states in the gap between the $n=\numlist{3;4}$ bands at $B=\SIlist{4;8}{\tesla}$, \cf Fig.~\ref{fig:bulkproperties}(c).
We confirm these predictions in Fig.~\ref{fig:edgestates} by numerically calculating the edge states supported by a broad ribbon (20 unit cells wide) extended along $x$ with the particular edge termination of Fig.~\ref{fig:edgestates}(a). The bulk states are folded into the projected 1D BZ, $k_x\in(-\pi/a,\pi/a)$, see Fig.~\ref{fig:edgestates}(b), due to breaking of Bloch periodicity along $y$. Additionally, edge states emerge: they are identified and post-selected from the ribbon-spectrum by their edge confinement and bulk-gap habitation (in emulating single-boundary physics, edge states localized on the bottom ribbon edge are omitted). The resulting edge-dispersion is shown in Fig.~\ref{fig:edgestates}(c) for $B=\SIlist{0;4;8}{\tesla}$.
At $B=0$, all edge states are non-topological; states at $\pm\mathbf{k}$ travel in opposite directions and edge connections between bulk bands are trivial. For $B\neq0$, topological one-way edge states appear in the bandgaps, consistent with the obtained gap Chern numbers. They connect upper and lower bulk bands, occasionally by circling the 1D BZ, separated by nontrivial $\bar{C}^{(n)}\neq0$ gaps. The edge states propagate to the right, consistent with the sign (chirality) of $ \bar{C}^{(n)}\neq0$. They are topologically protected from backscattering only in the complete bandgap: above it, any defect may scatter them to either bulk or counterpropagating edge states.
The low-frequency $\bar{C}^{(1)} = -1$ gap hosts edge states entirely analogous to the edge magnetoplasmons of the uniform sheet -- an edge-state mirror of the bulk dispersion-agreement (${\appropto} \sqrt{k}$) between the $n=1$ band and the uniform sheet.
In contrast, the high-frequency (${\approx} \SI{15}{THz}$) edge state in the $\bar{C}^{(2)} = -1$ gap result directly from band-engineering, and is a qualitatively new type of edge magnetoplasmon.
Finally, a multimode triple of edge states appear in the $\bar{C}^{(4)} = -3$ gap. Though the gap is comparatively small, it can be widened by tuning $a/d$.
Figure~\ref{fig:edgestates}(d) illustrates the sharp spatial Bloch mode confinement of the edge states, $|\phi_{nk_x}(\rv)|$, for a few select $n$ and $k_x$ at $B=\SI{8}{\tesla}$. The degree of confinement correlates positively with the size of the topological bandgap, \ie implicitly with $B$, paralleling the uniform 2D electron gas~\cite{Volkov:1988}.

The edge states can be efficiently excited by nearby point sources, as demonstrated in Fig.~\ref{fig:dipoleexcitation}: a $y$-polarized dipole near the edge, emitting in the gap-center ($\SI{14.6}{THz}$) of the $n=\numlist{1;2}$ bands, excites the edge plasmon at $k_x=0$ (for computational details, see SM). In the absence of intrinsic material loss, the edge state propagates unidirectionally to the right with constant amplitude as seen in Figs.~\ref{fig:dipoleexcitation}(b-d). Topological protection ensures that even structural defects, such as the sharp trench in Fig.~\ref{fig:dipoleexcitation}(d), are traversed without backscattering. The increased edge-confinement with mounting magnetic field is exemplified by Figs.~\ref{fig:dipoleexcitation}(b-c).

The edge state's topological nature does not shield from intrinsic material or radiation loss. While the latter is negligible, owing to the strongly localized and electrostatic nature of graphene plasmons [\cf the nearly vertical light cone in Fig.~\ref{fig:bulkproperties}(c)], the former can be appreciable, as in all plasmonic systems. Finite relaxation $\gamma$ is readily incorporated in Eqs.~\eqref{eqs:governing} by the substitution $\omega_\nu\rightarrow \omega_\nu+\Ii\gamma$. This introduces an imaginary spectral component, $\Im\omega_\nu \simeq -\tfrac{1}{2}\gamma(1 + \xi_\nu\omegac/\!\Re \omega_\nu^0)$ for $\gamma\ll\Re\omega_\nu^0$. This impacts the propagation of edge states in two aspects: first, it blurs the gap region, allowing small but finite loss-induced coupling between edge and bulk states (see SM); second, states exhibit a finite life-time, or, equivalently, finite propagation length ${\appropto}\,1/\gamma$.

In conclusion, we have demonstrated the band topology of 2D plasmons in periodically patterned graphene under a $\mathcal{T}$-breaking magnetic field. Multiple sets of topologically protected one-way edge plasmons corresponding to nontrivial gap Chern numbers are discovered. Their operating frequencies can be as high as tens of THz, \ie in the far-infrared regime. They can be experimentally verified by terahertz near-field imaging~\cite{Chen:2012,Fei:2012} and Fourier transform infrared spectroscopy \cite{Skirlo:2015}.
Our findings suggests a new direction in the synthesis of high-frequency $\mathcal{T}$-broken topological bosonic phases, and can be directly extended to non-magnetic schemes based on valley polarization~\cite{Kumar:2016,Song:2016}.

\FloatBarrier %Force execution of floats before acknowledgements and reference list

%----------------------------
%----- ACKNOWLEDGEMENTS -----
%----------------------------

\vskip 1.5em
\begin{acknowledgments}
DJ, NXF, and XZ acknowledge financial support by the NSF (Grant No.~CMMI-1120724) and AFOSR MURI (Award No.~FA9550-12-1-0488).
TC acknowledges financial support from Villum Fonden and the Danish Council for Independent Research (Grant No.~DFF--6108-00667).
MS was supported in part by the Army Research Office through the Institute for Soldier Nanotechnologies (Contract No.~W911NF-13-D-0001).
LL was supported in part by the National Key Research and Development Program of China (Grant No.~2016YFA0302400) and in part by the National Thousand-Young-Talents Program of China.
\end{acknowledgments}

%--------------------------------------------------------------------------------------------------------

%----------------------
%----- REFERENCES -----
%----------------------

\bibliographystyle{apsrev4-1}

\end{document}